\documentclass[sn-mathphys,Numbered]{sn-jnl}

\usepackage{graphicx}%
\usepackage{multirow}%
\usepackage{amsmath,amssymb,amsfonts}%
\usepackage{amsthm}%
\usepackage{mathrsfs}%
\usepackage[title]{appendix}%
\usepackage{xcolor}%
\usepackage{textcomp}%
\usepackage{manyfoot}%
\usepackage{booktabs}%
\usepackage{algorithm}%
\usepackage{algorithmicx}%
\usepackage{algpseudocode}%
\usepackage{listings}%
\usepackage{aas_macros}

\newcommand{\be}{\begin{equation}}
\newcommand{\ee}{\end{equation}}

\newcommand{\unit}[1]{\mathrm{#1}}

\newcommand{\Msun}{M_{\odot}}

\newcommand{\msunpyr}{\Msun\unit{yr}^{-1}}

\newcommand{\JADEStwelve}{JADES-GS-z12-0~}
\newcommand{\JADESeleven}{JADES-GS-z11-0~}

\newcommand{\JADESzthirteen}{JADES-GS-z13-0~}

\theoremstyle{thmstyleone}%
\theoremstyle{thmstyletwo}%

\theoremstyle{thmstylethree}%

\begin{document}

\title[UHZ1 as a SMDSs remnant]{UHZ1 and the other three most distant quasars observed: possible evidence for Supermassive Dark Stars}


\author*[1]{\fnm{Cosmin} \sur{Ilie}}\email{cilie@colgate.edu}

\author[2,3,4]{\fnm{Katherine} \sur{Freese}}\email{ktfreese@utexas.edu}

\author[5,6]{\fnm{Andreea} \sur{Petric}}\email{apetric@stsci.edu}

\author[7]{\fnm{Jillian} \sur{Paulin}}\email{jpaulin@sas.upenn.edu}


\affil*[1]{\orgdiv{Department of Physics and Astronomy}, \orgname{Colgate University},\\ \orgaddress{\street{13 Oak Dr}, \city{Hamilton}, \state{NY}, \postcode{13346}, \country{U.S.A}}}

\affil[2]{\orgdiv{Weinberg Institute for Theoretical Physics, Texas Center for Cosmology and Astroparticle Physics}, \orgname{University of Texas}, \\\orgaddress{\street{2515 Speedway}, \city{Austin}, \state{TX},\postcode{78712}, \country{U.S.A}}}

\affil[3]{\orgdiv{The Oskar Klein Centre, Department of Physics}, \orgname{Stockholm University}, \city{Stockholm}, \country{Sweden}}

\affil[4]{\orgdiv{Nordic Institute for Theoretical Physics (NORDITA)}, \orgname{AlbaNova University Centre}, \orgaddress{\street{Hannes Alfv\'{e}ns v\"{a}g 12}, \city{Stockholm}, \postcode{SE-106 91}, \country{Sweden}}}

\affil[5]{\orgname{Space Telescope Science Institute (STScI)},\orgaddress{\street{ 3700 San Martin Dr}, \city{Baltimore}, \state{MD}, \postcode{21218}, \country{U.S.A}}}

\affil[6]{\orgname{William H. Miller III Department of Physics and Astronomy, Johns Hopkins University}, \orgaddress{\city{Baltimore}, \state{MD}, \postcode{21218}, \country{U.S.A}}}

\affil[7]{\orgdiv{Department of Physics and Astronomy}, \orgname{University of Pennsylvania},\orgaddress{\street{ 209 S 33rd St}, \city{Philadelphia}, \state{PA}, \postcode{19104}, \country{U.S.A}}}


\abstract{The James Webb Space Telescope (JWST) has recently uncovered a new record-breaking quasar, UHZ1, at a redshift of $z\sim10$. This discovery continues JWST's trend of confronting the expectations from the standard $\Lambda$CDM model of cosmology with challenges. Namely, too many very massive galaxies and quasars have been observed at very high redshifts, when the universe was only a few hundred million years old. We have previously shown that Supermassive Dark Stars (SMDSs) may offer a solution to this puzzle. These fascinating objects would be the first stars in the universe, growing to be $\sim 10^5-10^7 \Msun$ and shining as bright as $10^9$ suns. Unlike Population III stars (the major alternative proposed model for the first stars in the universe, which would also have zero metallicity and would be powered by nuclear fusion), SMDSs would be powered by dark matter heating (e.g. dark matter annihilation) and would be comparatively cooler. At the ends of their lives (when they run out of dark matter fuel), SMDSs would directly collapse into black holes, thus providing possible seeds for the first quasars. Previous papers have shown that to form at $z\sim10$, UHZ1  would require an incredibly massive seed ($\sim 10^4 -10^5 \Msun$), which was assumed to be a Direct Collapse Black Hole (DCBH). In this paper, we demonstrate that Supermassive Dark Stars (SMDSs) offer an equally valid solution to the mystery of the first quasars, by examining the four most distant known quasars: UHZ1, J0313-1806, J1342+0928, and J1007+2115, with particular emphasis on UHZ1.}




\maketitle

\section{Introduction}\label{sec1}

The most distant observed quasars in the universe~\citep[see Refs.~][for statistical studies]{Wang:2019,Inayoshi:2020,lupi:2021} pose significant challenges to the standard theory of the formation of the first stars in the Universe. Namely, there is simply not enough time for the Supermassive Black Holes (SMBHs) powering them to grow to the masses they have (inferred from X-ray data) if one assumes they started from Black Holes (BHs) seeded by much lighter nuclear fusion--powered, zero metallicity stars. Those objects, also called Population~III (or Pop~III) stars, can grow to be as massive as $\sim100\Msun$.\footnote{For the interested reader  on Pop~III stars see:\cite{Abel:2001,Barkana:2000,Bromm:2003,Yoshida:2006,OShea:2007,Yoshida:2008,Bromm:2009}} If one insists on this kind of ``low seed" BHs, unrealistic sustained super-Eddington accretion rates are required. The most striking example of this puzzle is that of UHZ1, the most distant quasar observed to date. This extremely distant object ($z\simeq 10.2$) harbors a SMBH with   a mass of $10^7-10^8\Msun$, as inferred from Chandra X-ray data in Ref.~\cite{Bogdan:2023UHZ1}. Therefore the need for heavy, or even supermassive Black Hole seeds becomes evident. Refs.\cite{Bogdan:2023UHZ1,Natarajan:2023UHZ1} consider UHZ1 as evidence for the Direct Collapse Black Hole (DCBH) scenario~\citep[e.g.][]{Loeb:1994wv,Belgman:2006,Lodato:2006hw,Natarajan:2017,barrow:2018,Whalen:2020}. For a recent review of DCBHs see Ref.~\cite{Inayoshi:2020}. In this paper we point out an alternative interpretation: the mystery posed by the SMBHs powering the most distant four quasars (UHZ1, J0313-1806, J1342+0928, and J1007+2115) can be equally well solved by SMBHs seeded by Supermassive Dark Stars. 

The effects of Dark Matter (DM) heating on the formation of the first stars in the universe were first analyzed in Ref.~\cite{Spolyar:2008dark}, who found the conditions necessary for the formation of stars powered by annihilations of Weakly Interacting Massive Particles (WIMPs); later Ref.~\cite{Wu:2022SIDMDS} found that some types of Self Interacting Dark Matter (SIDM) could be the power source as well. Simply put, whenever zero metallicity molecular clouds collapse in regions of sufficiently high DM density a new object can form: a star powered exclusively by Dark Matter rather than nuclear fusion, i.e. a Dark Star (DS). Those conditions are typically met at the centers of high redshift ($z\sim 10-30$) DM minihalos, where the first stars in the Universe are expected to form. While Dark Stars are composed mainly of H and He, with less than $1\%$ in their mass carried by DM particles, Dark Stars are kept in hydrostatic and thermal equilibrium by Dark Matter heating alone.  The equilibrium structure of Dark Stars (DSs) can be well approximated using polytropes of variable index~\citep{Freese:2008ds,Rindler-Daller:2015SMDS}, ranging from $n=1.5$ (fully convective when the star is born, i.e. $M\sim 1\Msun$) to $n=3$ (radiation pressure dominated, when the star becomes more massive than $\sim 100\Msun$). Ref.~\cite{Rindler-Daller:2014uja} used the stellar evolution code MESA to obtain detailed models (that essentially confirmed the results using polytropes). In Ref.~\cite{Freese:2010smds} two mechanisms were identified (DM Capture and extended Adiabatic Contraction) via which DSs can grow to become supermassive ($M\sim 10^6\Msun$) and shine as bright as a billion suns or even more. Such bright objects should be easily detectable with the James Webb Space Telescope (JWST)~\citep{Ilie:2012} and the upcoming Roman Space Telescope~\citep{Zhang:2022}. As long as they are fueled by DM and as long as there is baryonic material surrounding them, DSs will continue to grow. Once the DM fuel is depleted, or if the SMDS is shifted from the high DM density location at the center of the host DM halo (by mergers for example) it will collapse to a SMBH. Thus Dark Stars provide a natural, often ignored, path to explaining the ``heavy seeds'' BHs required to explain the most distant quasars X-ray data. 

Dark Stars could also be part of the solution to a recent mystery in Astronomy. Within two years of becoming operational, the JWST has ushered in a revolution in our understanding of the cosmic dawn era. It keeps discovering very bright, compact sources during an epoch where, according to simulations, the first galaxies should have barely begun to assemble. This is the so-called ``too many too massive too soon'' problem posed by the JWST data~\citep[e.g.][]{GLASSz13, Maisies:2022, z16.CEERS93316:2022, z17.Schrodinger:2022,JADES:2022a,JADES:2022b,Labbe:2022}. The most striking and popularized examples of this puzzle are the six galaxies found by Ref.~\cite{Labbe:2022}. Those enormously bright sources emitted light when the universe's age was merely 500-700 Myr. If interpreted as galaxies, those sources would be more massive than ten billion suns. Neither numerical simulations, nor observations before JWST hinted to such monster galaxies existing so early. For this reason, those galaxies were sometimes referred to as cosmology (or $\Lambda$-CDM) breakers.\footnote{However, as shown by Ref.~\cite{Sabti:2023}, any $\Lambda$-CDM modifications that successfully alleviate the tension between JWST data and simulations, would, necessarily have observable consequences that should have been detected by the Hubble Space Telescope. As such, it is unlikely that any changes to the standard $\Lambda$-CDM cosmology are a viable solution to this tension.} Taken at face value, the JWST data is telling us that many of the first ``galaxies'' were converting gas to stars at an incredibly high, almost $100\%$, efficiency~\citep{Boylan-Kolchin:2023}. Dark Stars provide a natural explanation for these bright objects, as their surface temperatures are too low to shut off accretion. In fact, this is the main reason they can grow to become supermassive. One SMDS would shine as brightly as an entire galaxy of early smaller fusion-powered stars.  As such, SMDSs can provide a natural solution to the  ``too many too massive too soon'' puzzle, in addition to ``heavy seeds'' for the BH mystery posed by the most distant quasars observed. 

The SMDS hypothesis is already consistent with data from JWST. In Ref.~\cite{Ilie:2023JADES}  the first three SMDSs candidates were identified: \JADESeleven, \JADEStwelve, \JADESzthirteen,  with the last one being the most distant spectroscopically confirmed object in the universe ($z_{spec}\simeq 13.2$). Each of those three JWST Advanced Deep Extragalactic Survey (JADES) objects is currently also consistent with a high redshift compact galaxy interpretation~\citep{JADES:2022a,JADES:2022b}. 
As JWST continues to discover more high redshift objects, some of them lensed to high magnification, it will become possible  for 
future spectroscopic studies with the low-resolution spectrograph on JWST's Mid-Infrared Instrument (MIRI)  and the Near Infrared Spectrograph (NIRSpec) to differentiate between SMDS and galaxy interpretations~\citep{Ilie:2023JADES}. In this paper, we will focus on Dark Stars as possible seeds for the SMBHs powering the four most distant quasars observed: UHZ1, J0313-1806, J1342+0928, and J1007+2115, with particular emphasis on UHZ1, a source at an extremely high $z\sim 10$.

\section{Two kinds of ``heavy seeds:" Supermassive Dark Stars and DCBHs}\label{sec:SMDSToSMBHs}

In this section we demonstrate that Supermassive Dark Stars provide a natural mechanism to generate such seeds efficiently. 
The most prominent alternative scenario proposed in the literature is that of  Direct Collapse Black Holes (DCBHs), which we briefly review here and contrast against the SMDS seed scenario. Following Ref.~\cite{Bogdan:2023UHZ1}, our focus throughout this paper will be UHZ1, a recently discovered galaxy at $z\sim 10$, harboring a very bright quasar ($L_{bolo}\sim 5\times 10^{45}~\unit{erg\, s^{-1}}$). This object is the most striking example of the puzzle described above, in view of its extreme distance ($z\sim 10$) and inferred black hole mass ($M_{BH}\sim 10^7\Msun$). Refs.~\cite{Bogdan:2023UHZ1,Natarajan:2023UHZ1} conclude that UHZ1 is the best evidence so far for the DCBH scenario. In this work, we propose an alternative interpretation, by showing that the SMBH powering the X-ray observed spectrum of UHZ1 could be the Black Hole remnant of a Supermassive Dark Star.

\begin{figure}[!htb]
\includegraphics[width=0.9\textwidth]{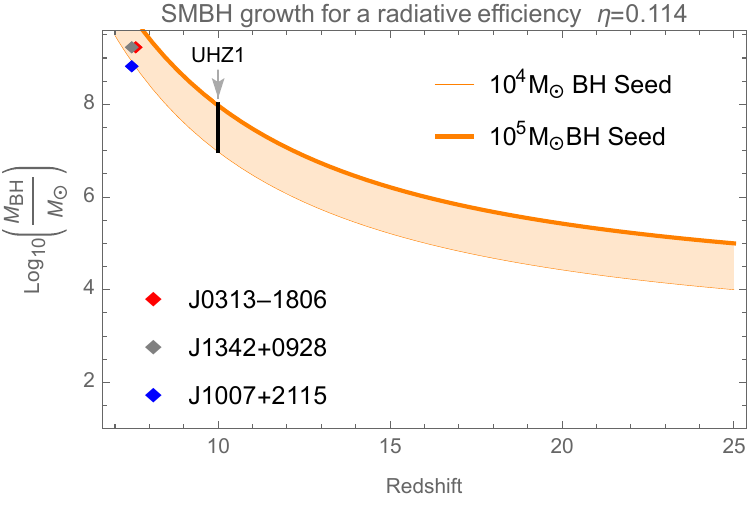}
\caption{BHs with masses between $10^4$ and $10^5 \Msun$, generated at $z\simeq 25$ and growing at the Eddington rate (tan shaded band), can explain the  mass of UHZ1 (Solid black line) and the three previously known highest redshift quasars (denoted by diamond symbols at $z\sim 7.5$). We found that a relatively large radiative efficiency of the accretion process ($\eta\simeq 0.114$) is needed to reproduce the growth curves from Fig.~4 of Ref.~\cite{Bogdan:2023UHZ1} (shaded tan band in this figure).}
\label{fig:DCBHGrowth}
\end{figure}

We start by briefly summarizing the main evidence for the ``heavy seed'' hypothesis as an explanation for UHZ1~\citep{Bogdan:2023UHZ1,Natarajan:2023UHZ1}. In Fig.~\ref{fig:DCBHGrowth} we plot, following Ref.~\cite{Bogdan:2023UHZ1}, the mass of hypothetical supermassive black holes that have been seeded at $z\simeq 25$ with masses in the expected DCBH initial mass range: $10^{4}-10^{5} \Msun$. We note here that SMDSs could collapse to BHs of the same mass range at the same redshift. Therefore, we argue that in fact UHZ1 is in fact the best evidence so far for the need of ``heavy seed'' Black Holes at high redshifts, rather than for the DCBH scenario itself.  

In order to generate the growth curves (i.e. the orange band in the figure) we solve:

\be\label{Eq:EddGrowth}
\dot{M}=4\times 10^{-8}\times \left(\frac{0.057}{\eta}\right) \times\frac{M}{\Msun}\times\Msun\unit{yr^{-1}}.
\ee
Eq.~\ref{Eq:EddGrowth} represents the growth rate of a quasar that shines due to accretion at the Eddington limit ($L_E=1.3\times 10^{38}~\unit{erg\,s^{-1}}$). Throughout we will denote by $\eta$ the efficiency of the accretion process:

$$
L=\eta\dot{M}c^2
$$
Typical expected values for this parameter are $\eta\sim 0.1$, with the fiducial value of $0.057$ being estimated including general relativistic effects~\citep[see Eq.7.39 of Ref.][]{Maoz:2016}. Note that if we were to use $\eta=0.057$ in Fig.~\ref{fig:DCBHGrowth}, the curves would have overshot the mass of UHZ1 and the $z\simeq 7.5$ quasars by orders of magnitude. We find that for $\eta\simeq 0.114$ we are able to reproduce the growth rates of Fig.~4 from Ref.~\cite{Bogdan:2023UHZ1}. Since Eddington accretion leads to an  exponential mass growth, even a small change in $\eta$ would imply a significant change in the estimated BH mass at $z\sim 10$ or below. This implies quite a large degree of fine-tuning is necessary to explain UHZ1 and the other three $z\simeq 7.5$ quasars in terms of DCBHs formed at $z\simeq 25$. However, since there is a wide range of possible formation redshifts ($z\in[15,30]$), this problem is alleviated, as the effects on the growth curves of lowering $z_{form}$ can be compensated by decreasing $\eta$.   

\begin{figure}[!htb]
\includegraphics[width=0.9\textwidth]{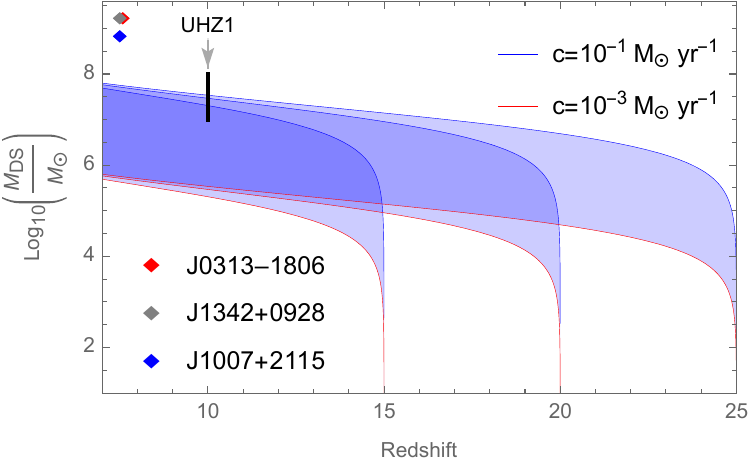}
\caption{Mass of Dark Stars as a function of redshift. For each formation redshift considered (15, 20, 25) the Dark Star initially has $M_{ini}\sim 1\Msun$ and grows via accretion at a constant rate, ranging between $c=10^{-1}$ (top/blue curves) and $c=10^{-3}\msunpyr$ (bottom/red curves). We also plot the UHZ1 mass range (black line at $z\simeq 10$, and the estimated mass of the other three highest redshift quasars (diamond symbols at $z\simeq 7.5$). Note that the growth of SMDSs quickly asymptotes to $M_{SMDS}\sim c\cdot t(z)$ regardless of the redshift of formation. In the upcoming Figures 3 and 4, we will show that SMDSs that collapse to SMBHs at $z_{BH} \sim 15-20$ can then grow via accretion to explain all four quasars shown here.}
\label{fig:SMDSGrowth}
\end{figure}

Dark Stars are expected to form in isolation at the centers of DM minihalos at redshifts $z_{form}\in[30,15]$. Their initial masses are  $\sim 1\Msun$ and they grow via accretion for as long as they are powered by DM annihilation and for as long as there is material to accrete~\citep{Freese:2010smds}. The accretion rate can be modeled with (constant) values typically ranging between $10^{-3}$ and $10^{-1} \msunpyr$. In Fig.~\ref{fig:SMDSGrowth} we plot the mass of Dark Stars as a function of redshift. To be concrete we assume they form at three different values of $z_{form}$: 25, 20, and 15. 
For each formation redshift the shaded blue region represents a range of possible masses of Dark Stars as a function of redshift, assuming they survive up to that point. We note that Dark Stars, if they survive by $z\sim 10$, could reach a mass $M_{SMDS}\gtrsim 10^{7}\Msun$, which is well within the estimated mass range of the BH within UHZ1. Once a Supermassive Dark Star runs out of fuel, it will immediately collapse into a Supermassive Black Hole that can subsequently grow via accretion or mergers. Thus, SMDSs can provide natural seeds for the SMBHs powering the most distant quasars: UHZ1, J0313–1806, J1342+0928,  J1007+2115, for which a low mass seed is highly unlikely~\citep{Bogdan:2023UHZ1}. In what follows we illustrate this point explicitly.  

\begin{figure}[!htb]
\includegraphics[width=0.9\textwidth]{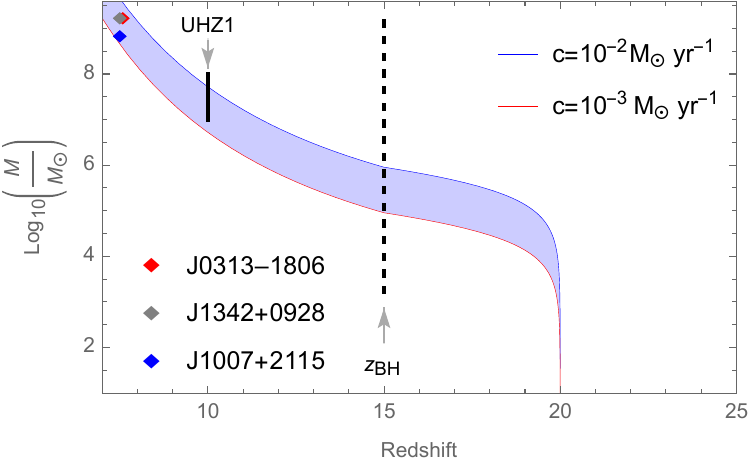}
\caption{Possible solution of the puzzle presented by UHZ1, J0313–1806, J1342+0928, and J1007+2115 in terms of Supermassive Black Holes seeded by Dark Stars. We consider a Dark Star-forming at $z_{form}$  and growing via accretion at a constant rate until it runs out of DM fuel and thus dies as a SMBH, at a redshift labeled by $z_{BH}$. In the plot the Dark Star phase evolution is depicted by the shaded blue region to the right of the vertical dashed line at $z=z_{BH}$. Subsequently, the emerging SMBH is assumed to grow at the Eddington accretion limit (blue-shaded region to the left of $z=z_{BH}$). To be concrete we chose $z_{form}=20$, $z_{BH}=15$, and the Dark Star mass accretion rate ranging in the conservative range $c\in[10^{-3},10^{-2}]\msunpyr$.}
\label{fig:SMDSSMBH}
\end{figure}

In Fig.~\ref{fig:SMDSSMBH} we demonstrate that Dark Stars can provide a solution to the mystery presented by the enormous mass of four most distant observed quasars. The idea is simple: a dark star is born at a redshift $z_{form}$, and it subsequently grows at a constant accretion rate $c$ until it runs out of DM fuel (or is dislodged from the center of the DM halo via mergers)  and collapses to a Supermassive Black Hole, at a redshift $z_{BH}$. The SMBH seeded this way continues to grow via accretion, possibly attaining masses in excess of $10^9\Msun$ by $z\simeq 7.5$. For concreteness we chose the following values for our three parameters $z_{form}=20$, $z_{BH}=15$, and the accretion rate in the conservative range $c\in[10^{-3},10^{-2}]\msunpyr$. We find that the SMBHs seeded by this mechanism and growing at the Eddington rate~\footnote{To be consistent with the DCBH scenario discussed in Fig.~\ref{fig:DCBHGrowth} we chose the same value for the radiative efficiency: $\eta=0.114.$} can explain the large inferred mass for UHZ1 and the three other most distant quasars observed: J0313–1806, J1342+0928,  J1007+2115. Note that in addition to our choice for the values for $z_{form}$, $z_{BH}$ and $c$ made in Fig.~\ref{fig:SMDSSMBH} there is a wide range of values for those parameters that leads to the same predicted BH mass range at $z\lesssim 10$. 

\begin{figure}[!htb]
\includegraphics[width=0.48\textwidth]{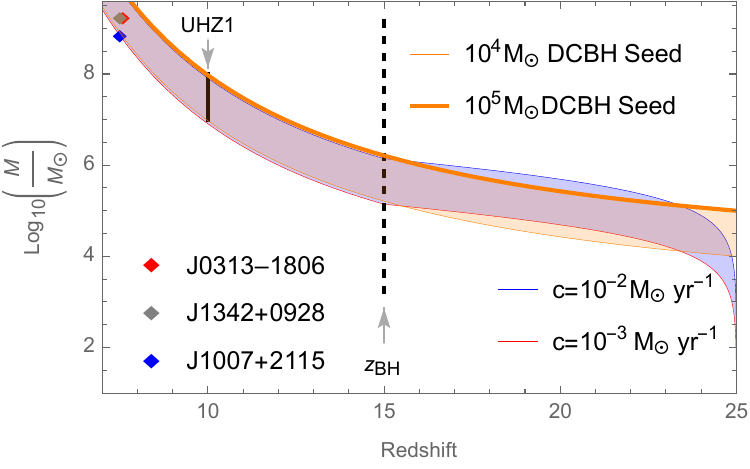}
\includegraphics[width=0.48\textwidth]{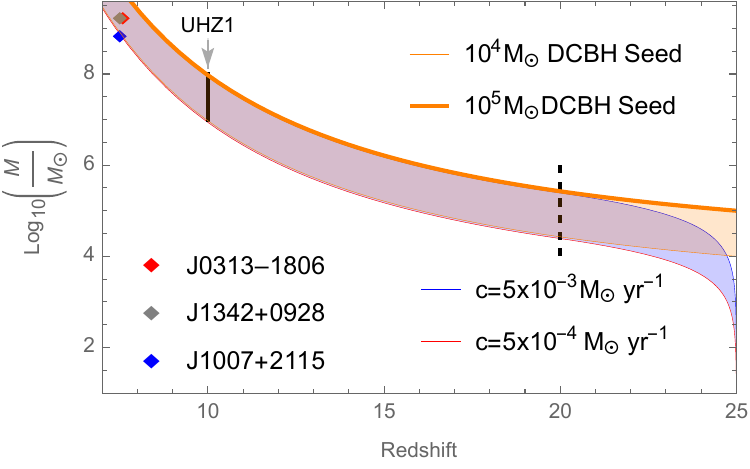}
\caption{In this figure we demonstrate that there is a wide range of parameters for which the Dark Stars solution (blue band)  to the high redshift SMBHs puzzle (such as UHZ1) is essentially degenerate to the DCBH solution (tan band, already discussed in Fig.~\ref{fig:DCBHGrowth}). For both the left and right panel we chose the formation redshift for the DS to be $z_{form}=25$, whereas the redshift at which the DS collapses to a BH ($z_{BH}$; denoted by dashed line) is different: 15 (left panel) vs. 20 (right panel). }
\label{fig:DCBHSMDSDeg}
\end{figure}

In Fig.~\ref{fig:DCBHSMDSDeg} we illustrate that the Dark Star seed solution is degenerate, in many cases, to the DCBH ``heavy seed'' scenario. Namely, there is a  wide range of parameters for which, by $z\lesssim 10$, the predicted mass of the SMDSs will be identical for both. For example, if one chose the same formation redshift: $z_{form}=25$, for both the DS and the DCBH, there are numerous ways to adjust the redshift at which the DS collapses ($z_{BH}$) and the rate at which it accretes ($c$) such that for  $z<z_{BH}$ both solutions are degenerate. For instance, for $z_{BH}=15$ we find that whenever $c\in[10^{-3},10^{-2}]\msunpyr$ the growth curves align perfectly after the collapse of the DS to a BH (see left panel of Fig.~\ref{fig:DCBHSMDSDeg}). Similarly, for $z_{BH}=20$, one needs only half the previous accretion rate: $c\in[5\times10^{-4},5\times10^{-3}]$. This degeneracy is further increased if we consider other possible formation redshifts for the DS (see for example Fig.~\ref{fig:SMDSSMBH}).

In conclusion, in this section we demonstrated that Dark Stars, in addition to DCBHs, are a plausible solution to the mystery of the origin of the SMBHs powering the most distant quasars in the universe. In what follows we discuss in more detail UHZ1, for which the JWST IR data hints at a large stellar population, with a mass approximately equal to that of the SMBH.

\section{Dark Star interpretation of UHZ1 data}

The JWST IR data suggests that UHZ1 harbors a significant stellar population. It is estimated that the stellar mass is roughly equal to the BH mass ($\sim 10^7\Msun$). This is actually a feature of the DCBH scenario~\citep{Bogdan:2023UHZ1,Natarajan:2023UHZ1}, since a large Pop~III stellar population is needed as a catalyst to the formation of a DCBH in a nearby satellite halo. In turn, there a variety of ways to explain the UHZ1 data using Dark Stars, the simplest of which is a SMBH SMDS system. The SMDS would be responsible for most of the IR observed flux, whereas the SMBH would explain the X-Ray observations. As shown in Sec.~\ref{sec:SMDSToSMBHs}, the SMBH could, in turn, have been produced by the collapse of a SMDS. However, in view of the extended nature of the UHZ1 system, as observed with NIRCam, this simple interpretation is highly unlikely. The second possible explanation of the data is in fact very similar to the one proposed by Refs.~\cite{Bogdan:2023UHZ1,Natarajan:2009}. Namely, the system is a galaxy that harbors a SMBH of $M_{SMBH}\sim 10^7\Msun$. In our scenario, what is different is the origin of the Supermassive Black Hole: a Dark Star or a DCBH. A natural question arises: can there be galaxies at the heart of which the remnants of Supermassive Dark Stars are found? Below we discuss a variety of such scenarios.  

\subsection{Mergers}\label{ssec:Mergers}

\begin{figure}[!htb]
\includegraphics[width=0.99\textwidth]{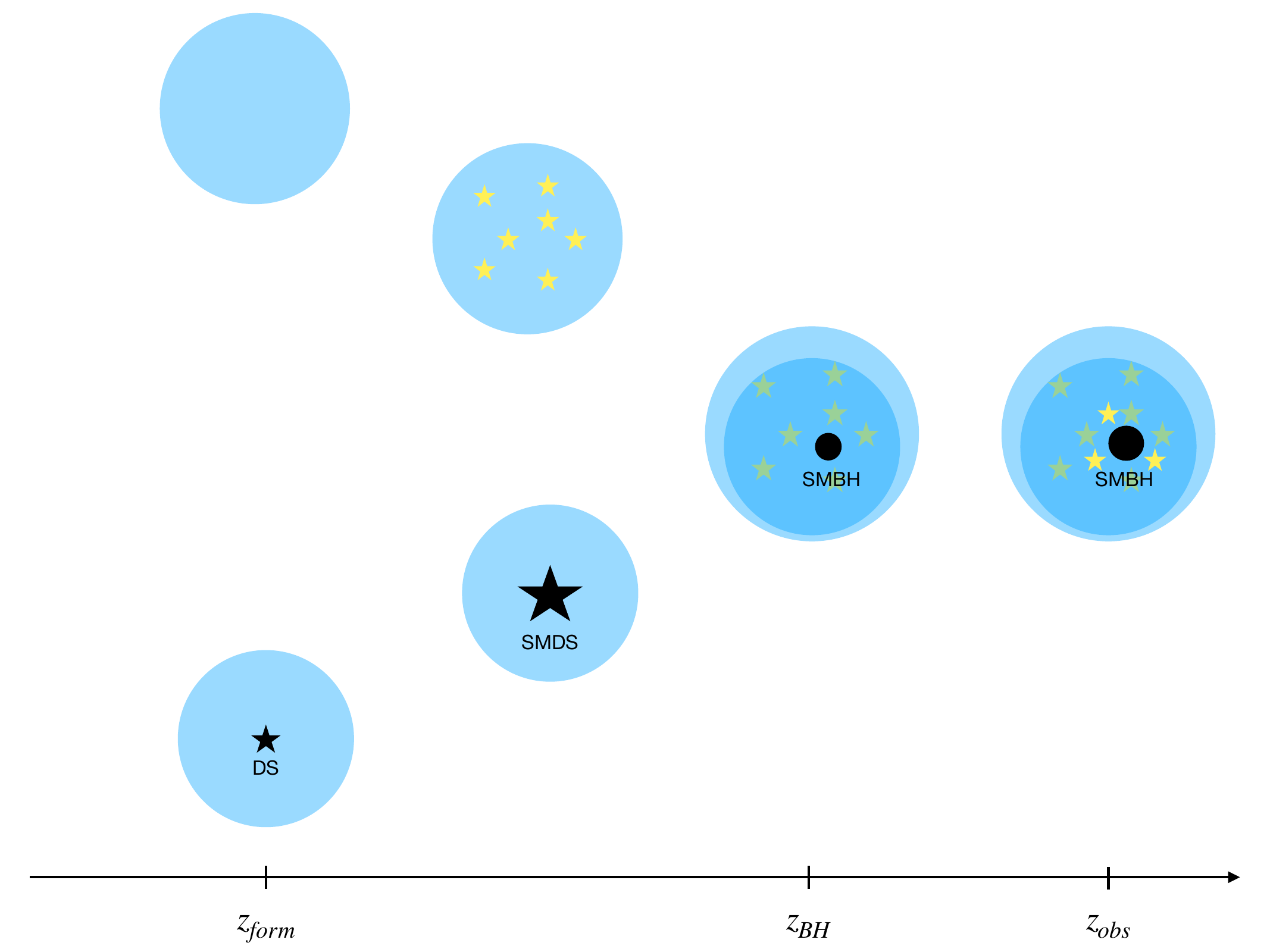}
\caption{Schematic representation of our interpretation of UHZ1 data in terms of the mergers of two DM halos. One of them hosts a Dark Star (formed at $z_{form}$) which can grow to become a SMDS. This halo merges with a second DM halo. In the figure this second halo hosts regular nuclear burning stars (represented by yellow stars); or this second halo could still be dark (and a starburst happens post-merger). As soon as it runs out of DM fuel the SMDS collapses to a SMBH (black dot). This collapse can be also triggered by the merger whenever the SMDS would be dislodged from high DM densities at the very center of the host halo. The baryonic reservoir in its vicinity is replenished by the merger, and the SMBH can accrete efficiently until observed. Moreover, ionizing radiation from the accreting SMBH triggers a starburst episode~\citep[e.g.][]{Silk:2005, Levin:2007MNRAS, carilli:2013, MF2023}(represented by bright yellow stars added for the depiction at $z_{obs}$) The growth of the DS$\to$SMDS$\to$SMBH has been modeled in Fig.~\ref{fig:DCBHSMDSDeg}, for example.}
\label{fig:Mergers}
\end{figure}

The morphology of UHZ1, as obtained from JWST's NIRCam data, is ``evocative of late stage mergers at low redshifts,'' as stated in Ref.~\cite{Bogdan:2023UHZ1}. As such, we consider the merger of two DM halos as the most plausible explanation of UHZ1. We discuss this scenario below, in the context of a Dark Star (DS) seed for the SMBH powering the X-ray spectra of UHZ1. We assume a DS forms in a high redshift DM halo. As discussed in Sec.~\ref{sec:SMDSToSMBHs}, any redshift for the formation of the DS ($z_{form}$) in the 25-15 range would work very well, with slightly lower or higher values possible. 
At a later stage, the halo hosting the (by now) Supermassive Dark Star begins to merge with another halo which could still be dark or could host regular nuclear fusion-powered stars. In some cases, the SMDS, due to its large mass, may remain at the center of the new larger object from the merger and continue to grow.  However, it is also possible that, as a result of the merger, some SMDSs get slightly dislodged from the high DM density at the center of the DM halos, a case we consider here. Given the high sensitivity (quadratic) of the DM heating inside the SMDS to the DM density, and the rapid decline of the DM density surrounding a DS with the distance from the center of the halo ($\sim r^{-1.9}$, as shown in Ref.~\cite{Spolyar:2008dark}) the merger of those two halos could then lead to the collapse of the SMDS to a SMBH, as it runs out of sufficient fuel to keep it in equilibrium. For such massive objects  once the fuel is reduced they collapse directly to Black Holes.\footnote{It is worth mentioning that even in the DCBH scenario, very often a supermassive star (SMS) is assumed/expected as a precursor to the ``Direct Collapse" SMBH, as explained in depth in Ref.~\cite{Inayoshi:2020}.} As a consequence of the merger, in addition to the collapse of the SMDS to a SMBH, an episode of rapid star formation, i.e. a starburst, is triggered in the vicinity of the SMBH~\citep[e.g.][]{Silk:2005, Levin:2007MNRAS, carilli:2013, heckman:2014, MF2023}. Moreover, the reservoir of baryons in the vicinity of the SMDS is replenished now via the merger. Thus, the SMBH can continue to grow at the Eddington rate, and be observed as the X-ray portion of the UHZ1 radiation. The stellar population embedded in those two merging halos is responsible for the majority of the IR observed data. In Fig.~\ref{fig:Mergers} we present a schematic representation of the scenario discussed above.   Or, there could already be one halo in which the SMDS has already collapsed to a SMBH, and that halo then merges with another. The merger then creates a starburst in addition to the SMBH that was already there.

\subsection{Same DM Halo Host}\label{ssec:OtherScenarios}

Would it be possible that all the stars within UHZ1 (including the Supermassive Dark Star we assume as the seed for the SMBH) have formed within the same DM halo? While highly unlikely (in view of the morphology of UHZ1), we briefly discuss this scenario below, as it might be relevant for other future observed high redshift quasars for which there is also a confirmed stellar counterpart.

Photons in the the Lyman-Werner (LW) band ($11.2-13.6$~eV) can photo-dissociate molecular H$_2$. Since molecular hydrogen is the key coolant mechanism available in zero metallicity pristine clouds that form the first stars, its destruction usually quenches formation rates significantly. This problem was initially discussed in Ref.~\cite{Omukai:1999}, who show that a Pop~III O5 type star, or hotter ($T_{eff}\gtrsim 54,000$~K), with mass $M\gtrsim 40\Msun$ can inhibit star formation in a region surrounding it as large as $r_{sh}\sim 1$kpc. Compared to the size of a high redshift ($z\sim 20$) DM halo, 1 kpc can be significant, or sometimes even exceed the size (as estimated by the virial radius) of lower mass DM minihalos. This would imply that star formation is almost completely (if not totally) quenched in any redshift DM halo, as soon as the first massive Pop~III star(s) are born within it and if there are no other efficient cooling mechanisms besides molecular hydrogen. 

However, the calculations done in Ref.~\cite{Omukai:1999} assume (among other things) that: i) the number density of molecular hydrogen quickly reaches a constant value controlled by the equilibrium between the dissociation and formation rates of H$_2$ within the vicinity of the star, and ii) that the baryonic number density is a constant throughout the DM halo. Neither of those assumptions represents a real situation. As discussed in Ref.~\cite{Glover:2000}, accounting for the time required for the LW flux created by a massive O5 star to photodissociate H$_2$ within its surroundings allows for star formation to continue efficiently in the most massive H$_2$ cooling DM halos ($M\sim 10^7\Msun$) or in the densest baryonic clumps (independent of the host halo mass). Moreover, if assumption (ii) is replaced with a commonly used isothermal sphere profile for the gas density surrounding the star (or any other realistic profile) the estimates of the radius of influence $r_{sh}$ would be significantly reduced. 

Dark Stars are inherently cooler than O5 stars, so their radius of influence would be significantly smaller. We find that a SMDS of $\sim 10^4\Msun$ emits the same LW flux as the $40\Msun$ Pop~III star considered by Refs.~\cite{Omukai:1999,Glover:2000}. Therefore it can take up to $10$ Myrs (assuming an accretion rate of $10^{-3}\msunpyr$) from the moment a Dark Star forms until it begins to significantly suppress the star formation in its host halo. As such, it is entirely possible that Pop~III stars can form from the collapse of gas $H_2$ clumps located the same parent DM halo as a Dark Star growing towards a SMDS.  Of course, once any of those Pop~III stars grows sufficiently massive the LW radiation will inhibit other stars from forming in their parent DM halo. 

The entire discussion above assumed that there are no other efficient cooling mechanisms besides molecular hydrogen. However, for DM halos with a virial temperature $T_{vir}\gtrsim 10^4$~K, atomic hydrogen cooling becomes efficient. As such, gas can cool irrespective of the flux of the LW radiation and regular star formation can continue indefinitely. Atomic Cooling Halos (ACHs) have a mass greater than $\sim 10^7\Msun$. The growth of SMDSs at the centers of atomic cooling haloes was studied using the MESA stellar evolution code in Ref.~\cite{Rindler-Daller:2014uja}.  Since the baryonic mass within the UHZ1 system must be greater than $10^7\Msun$, in the absence of halo mergers discussed in Section 3.1, the host DM halo must have been an atomic cooling halo. Thus it is plausible (although unlikely, in view of UHZ1's distorted morphology) that all the stars within UHZ1 (including the Supermassive Dark Star we assume as the seed for the SMBH) have formed within the same DM
halo. 

 \section{Discussion and Conclusions}
The ``low mass seed" scenarios in which a quasar at $z\gtrsim6$ originates from sustained Eddington-rate accretion onto an initially few stellar--mass black hole that is the remnant of a short-lived Pop~III star are challenging to reconcile with predictions of feedback which may easily unbound the gas and dust arresting the growth of the BH. Furthermore, growing small black holes through mergers is also challenging since the  rate is just too slow. Moreover, gravitational recoils during BH mergers can cause them to be ejected from the original halos \citep{Whalen:2004, Haiman:2001, Haiman:2004, Whalen:2020}. Finally, recent X-ray to NIR observations of high redshift QSOs suggest that the intrinsic photon index from 0.5 to 10 kEV cannot be explained by high Eddington ratios alone \citep{farina:2022}. These lines of evidence all favor a ``heavy seed" scenario, as further demonstrated by Ref.~\cite{Bogdan:2023UHZ1}. The IR luminosities, obscurations, and general observations of massive galaxies at the highest redshift suggest that the heavy seed black holes must grow without severely impacting star formation and dust production in the halos. Fueled by dark matter annihilations favored by the physical conditions in high redshift ($z\sim [10-25]$) dark matter halos, Dark Stars provide a solution to both enigmas of large galaxies and luminous QSOs observed by JWST at the highest redshifts. 

In this paper we demonstrate that even conservative accretion rates onto Dark Stars can evolve into the SMBH powering the four most distant observed quasars: UHZ1; J0313-1806, J1342+0928, and J1007+2115 (Fig.~\ref{fig:SMDSSMBH}). We also showed that a wide range of (i) formation redshifts for the Dark Star, (ii)  the accretion rate onto the DS, and (iii) the redshift when it collapses to a BH, produce the types of massive heavily obscured QSOs like UHZ1 (Figs.~\ref{fig:SMDSSMBH} and ~\ref{fig:DCBHSMDSDeg}). We further showed that subsequent mergers, similar to the late-stage galaxy mergers observed at low redshift, as suggested by the observed extended morphology of UHZ1, are compatible with a scenario in which the reservoir of baryons in the vicinity of SMDS is replenished via this merger (Figure \ref{fig:Mergers}). For UHZ1 this is the more likely SMDS interpretation, as discussed in Sec.~\ref{ssec:Mergers}. Finally, because Dark Stars are cooler than massive nuclear fusion powered O5 stars we predict less effective feedback on the surrounding medium permitting the growth of stars in the host galaxy, even in DM halos in which atomic hydrogen cooling is not efficient. Hence our second scenario (see Sec.~\ref{ssec:OtherScenarios}) also predicts a SMBH formed from an earlier SMDS as well as a population of nuclear-powered stars, in agreement with observations of UHZ1.

Regarding the alternative proposed  DCBH scenario for UHZ1 formed from supermassive stars, we point out that the abundance of halos at $z\gtrsim 10$ predicted to  host them is quite limited, according to simulations~\citep[e.g.][]{Natarajan:2017}. Thus, once a statistically significant sample of $z\gtrsim 10$ SMBHs is observed, one could place constraints on the DCBH scenario.  
In contrast, the alternative scenarios presented here, where SMBHs at high $z$ can be seeded by SMDSs, is more flexible in the sense that SMDS can form at a wide variety of redshifts and, a priori, the fraction of DM halos that can host a SMDSs could be quite large.

In summary, we argue that Supermassive Dark Stars can simultaneously offer a solution to the 
``too big - too soon" enigma for the massive high redshift galaxies as well as to
 explain UHZ1 and its high redshift luminous QSOs peers.

\bmhead{Acknowledgments}
K.F. is grateful for support from the Jeff and Gail Kodosky Endowed Chair in Physics  at the Univ. of Texas, Austin.   K.F.  acknowledges funding from the U.S. Department of Energy, Office of Science, Office of High Energy Physics program under Award Number DE-SC0022021. K.F. acknowledges support by the Vetenskapsradet (Swedish Research Council) through contract No. 638- 2013-8993 and the Oskar Klein Centre for Cosmoparticle Physics at Stockholm University. A.P. acknowledges STScI research funding D0101.90257. C.I. acknowledges funding from Colgate University via the Research Council.  

\bibliography{RefsDM,AOPref}

\end{document}